\documentstyle[12pt,graphicx,caption2]{article}

\begin{document}

\title{Models with non-Hermitian Hamiltonian and optical theorem}
\author{V.I.Nazaruk\\
Institute for Nuclear Research of RAS, 60th October\\
Anniversary Prospect 7a, 117312 Moscow, Russia.*}

\date{}
\maketitle
\bigskip

\begin{abstract}
The applicability of the optical theorem in the models with the non-Hermitian Hamiltonian is studied. By way of example we consider the $n\bar{n}$ transition in a medium followed by annihilation. It is shown that an application of optical theorem for the non-unitary $S$-matrix leads to the qualitative error in the result. The alternative model which is free from drawback given above is studied as well.
\end{abstract}

\vspace{5mm}

\vspace{5mm}

\vspace{1cm}

*E-mail: nazaruk@inr.ru

\newpage
\setcounter{equation}{0}
\section{Introduction}
The optical theorem should be applied for the unitary $S$-matrix since it follows from unitarity condition. It is frequently used for the non-Hermitian Hamiltonians as well. In this case the $S$-matrix should be unitarized. However, this requirement breaks down for a number of well-known models since the unitarization is the non-trivial aspect of the problem. 

In this paper the possible consequences are studied by the example of $n\bar{n}$ transitions [1-3] in a medium followed by annihilation
\begin{equation}
(n-\mbox{medium})\rightarrow (\bar{n}-\mbox{medium})\rightarrow (f-\mbox{medium}),
\end{equation}
where $f$ are the annihilation mesons. The matter is that in the standard calculations of this process (see [4-9], for example) the optical theorem is applied for the essentially non-unitary $S$-matrix. 

As an alternative to the standard calculation mentioned above we consider the simple model (see Fig. 1) with unitary $S$-matrix (later on reffered to as model with the Hermitian Hamiltonian). For both unitary and non-unitary models the solution in the analytical form is available which permits the generalization of the results.

It is shown that an application of optical theorem for the non-unitary $S$-matrix leads to the qualitative error in the result. The alternative approach gives the result which depends critically on the details of the model. Due to this the lower limit on the free-space $n\bar{n}$ oscillation time $\tau $ lies in the broad range $10^{16}\; {\rm yr}>\tau >1.2\cdot 10^{9}\; {\rm s}$. The explanation of the huge distinction between the values of $\tau $ ($10^{16}\; {\rm yr}$ and $1.2\cdot 10^{9}\; {\rm s}$) is the second purpose of this paper.

\section{Optical theorem and unitarity}
We recall that unitarity condition
\begin{equation}
(SS^+)_{fi}=\delta _{fi},
\end{equation}
$S=1+iT$, gives
\begin{equation}
2ImT_{ii}=\sum_{f\neq i}\mid T_{fi}\mid ^2.
\end{equation}
From this equation the optical theorem and expression for the decay width 
\begin{equation}
\Gamma _{opt}=\frac{1}{T_0}(1-\mid S_{ii}\mid ^2)\approx \frac{1}{T_0}2ImT_{ii}
\end{equation}
are obtained. Here $T_0$ is the normalization time, $T_0\rightarrow \infty $.
The non-unitarity of $S$-matrix implies that $(SS^+)_{fi}\neq \delta _{fi}$ or, what is the same
\begin{equation}
(SS^+)_{fi}=\delta _{fi}+\alpha _{fi},
\end{equation}
$\alpha _{fi}\neq 0$, resulting in 
\begin{equation}
2ImT_{ii}=\sum_{f\neq i}\mid T_{fi}\mid ^2-\alpha _{fi}\neq \sum_{f\neq i}\mid T_{fi}\mid ^2
\end{equation}		
since the value $\sum_{f\neq i}\mid T_{fi}\mid ^2$ can be very small. Instead of (2) we have (5) and eq. (3) and optical theorem are inapplicable. Also eq. (5) means the probability non-conservation: $\sum_{f}\mid S_{fi}\mid ^2\neq 1$. 

\section{Model with the non-Hermitian Hamiltonian}
Let us consider the process (1). The background potential of neutron-medium interaction $U_n$ is included in the neutron wave function: $n(x)=\Omega ^{-1/2}\exp (-i\epsilon t+i{\bf p}{\bf x})$, $\epsilon ={\bf p}^2/2m+U_n$.
In the standard calculation [4-9] the $\bar{n}$-medium interaction is described by optical potential (potential model). The interaction Hamiltonian is 
\begin{equation}
{\cal H}_I={\cal H}_{n\bar{n}}+{\cal H}_{opt},
\end{equation}
\begin{equation}
{\cal H}_{n\bar{n}}=\epsilon \bar{\Psi }_{\bar{n}}\Psi _n+H.c.,
\end{equation}
\begin{equation}
{\cal H}_{opt}=(U_{\bar{n}}-U_n)\bar{\Psi }_{\bar{n}}\Psi_
{\bar{n}}=(V-i\Gamma /2)\bar{\Psi }_{\bar{n}}\Psi_{\bar{n}}.
\end{equation}
Here ${\cal H}_{n\bar{n}}$ and ${\cal H}_{opt}$ are the Hamiltonians of $n\bar{n}$ conversion [4,5] and $\bar{n}$-medium interaction, respectively; $\epsilon $ is a small parameter with $\epsilon =1/\tau $, $U_{\bar{n}}$ is the antineutron optical potential, $\Gamma $ is the annihilation width of $\bar{n}$.  In eq. (9) we have put ${\rm Re} U_{\bar{n}}-U_n=V$, ${\rm Im} U_{\bar{n}}=-\Gamma /2$. 

The model can be realized by means of equations of motion [4-9], or diagram technique. The full in-medium antineutron propagator $G_m$ is
\begin{equation}
G_m=\frac{1}{\epsilon _{\bar{n}} -{\bf p}_{\bar{n}}^2/2m-U_{\bar{n}}+i0},
\end{equation}
${\bf p}_{\bar{n}}={\bf p}$, $\epsilon _{\bar{n}}=\epsilon $. The on-diagonal matrix element $T_{ii}$ is shown in Fig. 1c. For the total decay width $\Gamma _{opt}$ eq. (4) gives the well-known result [4-9]:
\begin{equation}
\Gamma _{opt}=-2{\rm Im}\epsilon G_m\epsilon =2\epsilon ^2\frac{\Gamma /2}{V^2+(\Gamma /2)^2}\approx 4\epsilon ^2/\Gamma .
\end{equation} 
The lower limit on the free-space $n\bar{n}$ oscillation time $\tau _{{\rm pot}}$ derived by means of eq. (11) is $\tau _{{\rm pot}}=2.36\cdot 10^{8}\; {\rm s}$ [7].  

However, the basic eq. (4) is inapplicable since $S$-matrix is non-unitary (see (9)). The model should be revised.

\section{Models with the Hermitian Hamiltonian}
In the one-particle model described above the total decay width can be obtained by means eq. (4) only. We calculate directly the off-diagonal matrix element in the framework of field-theoretical approach. The process model is shown in Fig. 1a. The interaction Hamiltonian is
\begin{equation}
{\cal H}_I={\cal H}_{n\bar{n}}+{\cal H},
\end{equation}
\begin{equation}
{\cal H}=V\bar{\Psi }_{\bar{n}}\Psi_{\bar{n}}+{\cal H}_a,
\end{equation}
Here ${\cal H}$ is the Hamiltonian of $\bar{n}$-medium interaction, ${\cal H}_a$ is the effective annihilation Hamiltonian in the second quantization representation, $V$ is the residual scalar field, ${\cal H}_{n\bar{n}}$ is given by (8).

\subsection{Model a}
We consider the model shown in Fig. 1a (model {\bf a}). The amplitude of antineutron annihilation in the medium $M_a$ is given by
\begin{equation}
<\!f0\!\mid T\exp (-i\int dx{\cal H}_a(x))-1\mid\!0\bar{n}_{p}\!>=
N(2\pi )^4\delta ^4(p_f-p_i)M_a.
\end{equation}
Here $\mid\!0\bar{n}_{p}\!>$ is the state of the medium containing the $\bar{n}$ with the 4-momentum $p=(\epsilon ,{\bf p})$; $<\!f\!\mid $ denotes the annihilation mesons, $N$ includes the normalization factors of the wave functions. $M_a$ includes the all orders in ${\cal H}_a$.		

\begin{figure}[h1]
  {\includegraphics[height=.25\textheight]{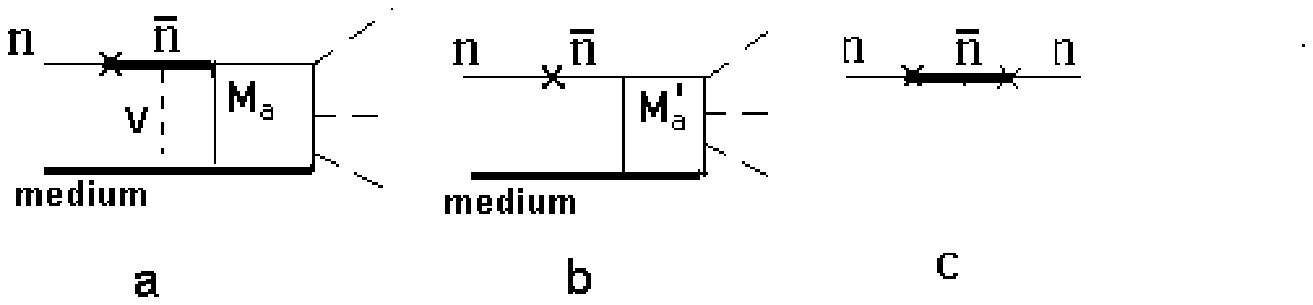}}
  \caption{{\bf a} $n\bar{n}$ transition in the medium followed by annihilation.  {\bf b} Same as {\bf a} but the annihilation amplitude is given by (21). The blocks $M_a$ and $M'_a$ involve the all orders in ${\cal H}_a$. {\bf c}  The on-diagonal matrix element $T_{ii}$ (see text)}
\end{figure}

In the lowest order in ${\cal H}_{n\bar{n}}$ the process amplitude is {\em uniquely} determined by the Hamiltonian ${\cal H}_I$:
\begin{equation}
M=\epsilon G_VM_a,
\end{equation}
where the antineutron Green function $G_V$ is
\begin{equation}
G_V=G+GVG+...=\frac{1}{(1/G)-V}=-\frac{1}{V},
\end{equation}
\begin{equation}
G=\frac{1}{\epsilon _{\bar{n}} -{\bf p}_{\bar{n}}^2/2m-U_n+i0} \sim \frac{1}{0},
\end{equation}
since ${\bf p}_{\bar{n}}={\bf p}$, $\epsilon _{\bar{n}}=\epsilon $. The Hamiltonian ${\cal H}_a$ acts in the block $M_a$ only and so $G_V$ is completely
determined by $V$.

For the total process width $\Gamma _a$ one obtains
\begin{equation}
\Gamma _a=N_1\int d\Phi \mid\!M\!\mid ^2=\frac{\epsilon ^2}{V^2}N_1\int d\Phi \mid\!M_a\!\mid ^2=\frac{\epsilon ^2}{V^2}\Gamma ,
\end{equation}
\begin{equation}
\Gamma =N_1\int d\Phi \mid\!M_a\!\mid ^2.
\end{equation}
The normalization multiplier $N_1$ is the same for $\Gamma _a$ and $\Gamma $.

The time-dependence is determined by the exponential decay law:
\begin{equation}
W_a(t)=1-e^{-\Gamma _at}\approx \frac{\epsilon ^2}{V^2}\Gamma t.
\end{equation}
The realistic set of parameters is $\Gamma =100$ MeV, and $V=10$ MeV. For the lower limit on the free-space $n\bar{n}$ oscillation time $\tau _a$ eq. (20) gives $\tau _a=5\tau _{{\rm pot}}=1.2\cdot 10^{9}\; {\rm s}$. 

\subsection{Model b}
We consider the model shown in Fig. 1b (model {\bf b}). If $M_a$ is determined by (14), the process amplitude (15) follows uniquely from (12), (13). On the other hand, for the one-step process of the antineutron annihilation in the medium $(\bar{n}-\mbox{medium})\rightarrow (\mbox{annihilation mesons}-\mbox{medium})$, the annihilation amplitude $M'_a$ can be defined through the Hamiltonian ${\cal H}$ and not ${\cal H}_a$:
\begin{equation}
<\!f0\!\mid T\exp (-i\int dx{\cal H}(x))-1\mid\!0\bar{n}_{p}\!>=
N(2\pi )^4\delta ^4(p_f-p_i) M'_a.
\end{equation}
$M'_a$ contains the all $\bar{n}$-medium interactions including antineutron rescattering in the initial state. In this case the amplitude of process (1) is
(see Fig. 1b)
\begin{equation}
M'=\epsilon GM'_a.
\end{equation}
The definition of annihilation amplitude through eq. (21) is natural since it corresponds to the observable values. There are many physical arguments in support of the model (22). However, this model contains infrared singularity
$M'\sim 1/0$ since $G\sim 1/0$. This problem has been considered in [8,9]. (In [8,9] only model {\bf b} has been studied.) For the purposes of this paper it is essential that model (22) gives linier $\Gamma $-dependence $\Gamma _b\sim \int d\Phi \mid\!M'\!\mid ^2\sim \Gamma $, as well as model (15). Consequently, the model with Hermitian Hamiltonian gives linear $\Gamma $-dependence at any definition of annihilation amplitude.

The lower limit obtained in the framework of the model {\bf b} is $\tau _b=10^{16}\; {\rm yr}$ [8,9]. If $V\rightarrow 0$, model {\bf a} converts to model {\bf b}. The huge distinction between the values of $\tau _a$ and $\tau _b$ is due to the $1/V^2$-dependence of (18).
 
\section{Discussion}
The $\Gamma $-dependence of the models with Hermitian and non-Hermitian Hamiltonians differs fundamentally: $\Gamma _{opt}\sim 1/\Gamma $, whereas $\Gamma _{a,b}\sim \Gamma $. At the same time the annihilation is the main effect which defines the process speed. One of two models is wrong. 

We assert that model with non-Hermitian Hamiltonian ${\cal H}_{opt}$ is wrong since (11) follows from (4) which is inapplicable for non-unitary $S$-matrix. 
Besides, ${\rm Im}T_{ii}$ is unknown (see below). Notice that the result (11) takes place in the all standard calculations [4-9] because they are based on the optical potential. 

We compare (18) and (11):
\begin{equation}
r=\frac{\Gamma _a}{\Gamma _{opt}}=\frac{\Gamma ^2}{4V^2}.
\end{equation}
For the parameters used in (20) ($\Gamma =100$ MeV and $V=10$ MeV) we have $r=25$. When $V=0$ as well as in the case of the model {\bf b}, eqs. (18) and (23) are invalid. However, in that event $r\gg 1$ as well [9].

On the other hand, for small $\Gamma $ eq. (11) {\em coincides} with (18):
\begin{equation}
2\epsilon ^2\frac{\Gamma /2}{V^2+(\Gamma /2)^2}\approx \frac{\epsilon ^2}{V^2}\Gamma .
\end{equation} 
This is because the Hamiltonian ${\cal H}_{opt}$ is {\em practically Hermitian} in this case. If $\Gamma \rightarrow 0$, the results of models with Hermitian and non-Hermitian Hamiltonians coincide. We would like to stress this fact. It can be considered as a test for the model given in sect. 4. Also we believe that the Hamiltonian ${\cal H}_{opt}$ describes correctly the $n\bar{n}$ transition with $\bar{n}$ in the final state $(n-\mbox{medium})\rightarrow (\bar{n}-\mbox{medium})$ since eq. (4) is not used in this case [10].

Consequently, for the non-unitary models the optical theorem can be used for the estimations if the absorption is small:
\begin{equation}
\mid {\rm Im} U_{\bar{n}}\mid \ll \mid {\rm Re} U_{\bar{n}}-U_n\mid.
\end{equation}
This is not the case for the $n\bar{n}$ transition in the nuclear matter. Because of this we performed the calculations in the framework of unitary 
models.

If the optical potential is used for the problems described by Schrodinger-type equation (optical model), the unitarization takes place [10]: the matrix elements and optical potential are fitted to $\bar{p}$-atom ($\pi ^-$-atom, $K^-$-atom) and low energy scattering data. However, the optical potential is the effective one. The $n\bar{n}$ transition is described by the system of coupled equations [5,8-10]. The corresponding $S$-matrix differs principally [10]. Even the physical meaning of ${\rm Im} U_{\bar{n}}$ is uncertain: one cannot get continuity equation from the system of coupled equations. {\em There are no experimental data and unitarization} in this case. Equations (3) and (4) are inapplicable. Besides, since ${\rm Im}T_{ii}$ is unknown, eq. (3) is inoperative in principle. (We also note that it is meaningless to impose the condition of probability conservation $\sum_{f}\mid S_{fi}\mid ^2=1$ since $S_{ii}$ is unknown.) At the same time, in the model described in sect. 3 eq. (4) is used, where $T_{ii}$ is unknown. The consequences are illustrated by eqs. (11), (18) and (23).

If the optical theorem is not applied, the range of applicability of optical potential is considerably wider. As a first approximation, it can be used in the calculation of the diagrams with $\bar{n}$ in the intermediate or final states [10]. For example, the channel $(n-\mbox{medium})\rightarrow (\bar{n}-\mbox{medium})$. In these cases the off-diagonal matrix elements are calculated directly without use of optical theorem.

In view of the uncertainty in the annihilation amplitude we cannot decide between models {\bf a} and {\bf b}. The same is true for the value of the antineutron self-energy $\Sigma =V$ in (16). These problems are general in the theory of reactions. However, in the problem under study the values of $\Gamma _a$ and $\tau _a$ are extremely sensitive to $V$ (see (18)). This is because the amplitude (22) is in the peculiar point $M'\sim 1/0$. The small change of $V$ affects the result vastly: $\Gamma _a\sim 1/V^2$. Owing to this $\tau _a$ and $\tau _b$ differ greatly. 

\section{Conclusion}
The main results are as follows. 

(a) For the non-unitary models the optical theorem and condition of probability conservation can be used for the estimations if the absorption is small. For the models with the essentially non-Hermitian Hamiltonians they can be applied only if $S$-matrix is unitarized. This is also true for the $K^0\bar{K}^0$ oscillations in particular.

(b) The huge distinction between the values of $\tau _a$ and $\tau _b$ stems from the fact that $\Gamma _a\sim 1/V^2$ and $V\rightarrow 0$. At present it is impossible to decide between models {\bf a} and {\bf b} as well as to determine the value of $V$ exactly. So the values $\tau _a=1.2\cdot 10^{9}\; {\rm s}$ and $\tau _b= 10^{16}$ yr are interpreted as the estimations from below (conservative limit) and from above, respectively. The realistic limit $\tau $ can be in the range $10^{16}\; {\rm yr}>\tau >1.2\cdot 10^{9}\; {\rm s}$. 
The estimation from below $\tau _a=1.2\cdot 10^{9}\; {\rm s}$ exceeds the restriction given by the Grenoble reactor experiment [11] by a factor of 14 and the lower limit given by potential model by a factor of 5. Further investigations are desirable.

\end{document}